\documentclass[aps,prl,twocolumn,showpacs,showkeys,superscriptaddress]{revtex4-1}
\usepackage{hyperref}
\usepackage[dvips]{graphicx}
\usepackage{amsmath}
\usepackage{amsfonts}
\usepackage[english]{babel}
\usepackage{bm}    

\newcommand{\be}{\begin{equation}}
\newcommand{\ee}{\end{equation}}

\newcommand{\relaxx}{\mathcal{R}}

\newcommand{\ka}{\bm{k}}
\newcommand{\pa}{\bm{p}}
\newcommand{\qa}{\bm{q}}
\newcommand{\va}{\bm{v}}
\newcommand{\ma}{\bm{m}}

\newcommand{\slz}{\mathrm{SL}(2,\mathbb{Z})}

\newcommand{\mysection}[1]{\vspace{2mm} \noindent{\sc #1\,:\;}}

\begin{document}

\title{Conservation laws for strings in the Abelian Sandpile Model}

\author{Sergio Caracciolo}
\affiliation{Universit\`a degli Studi di Milano -- Dipartimento di Fisica
      and INFN, via G.~Celoria 16, 20133 Milano, Italy}
\author{Guglielmo Paoletti}
\affiliation{Universit\`a  di Pisa -- Dipartimento di Fisica
      and INFN, largo B.~Pontecorvo 3, 56127 Pisa, Italy}
\author{Andrea Sportiello}
\setcounter{affil}{1}

\date{\today}

\begin{abstract}
\noindent
The Abelian Sandpile generates complex and beautiful patterns and
seems to display allometry.  On the plane, beyond \emph{patches},
patterns periodic in both dimensions, we remark the presence of
structures periodic in one dimension, that we call \emph{strings}. We
classify completely their constituents in terms of their principal
periodic vector $\ka$, that we call \emph{momentum}. We derive a
simple relation between the momentum of a string and its density of
particles, $E$, which is reminiscent of a dispersion relation,
$E=|\ka|^2$. Strings interact: they can merge and split and within
these processes momentum is conserved, $\sum_a \ka_a = {\bf 0}$. We
reveal the role of the modular group $\slz$ behind these laws.

\end{abstract}



\maketitle

\mysection{Introduction}%
Since the appearance of the masterpiece by D'Arcy
Thompson~\cite{Thompson}, there have been many attempts to understand
the complexity and variety of shapes appearing in Nature at
macroscopic scales, in terms of the fundamental laws which govern the
dynamics at microscopic level. Because of the second law of
thermodynamics, the necessary Self-Organization can be studied only in
non-equilibrium statistical mechanics.

In the context of a continuous evolution in a differential manifold,
the definition of a shape implies a boundary and thus a
discontinuity. This explains why catastrophe theory, the mathematical
treatment of continuous action producing a discontinuous result, has
been developed in strict connection to the problem of 
{\em Morphogenesis}~\cite{Thom}.
More quantitative results have been obtained by the introduction of
stochasticity, as for example in the diffusion-limited
aggregation~\cite{DLA}, where self-similar patterns with fractal
scaling dimension~\cite{Mandelbrot} emerge which suggest a relation
with scaling studies in non-equilibrium.

Cellular automata, that is, dynamical systems with discretized time,
space  and internal states, were originally introduced
by Ulam and von Neumann in the 1940s, and then commonly used as a
simplified description of phenomena like crystal growth, Navier-Stokes
equations and transport processes~\cite{CD}. They often exhibit intriguing
patterns~\cite{Wolfram}, and, in this regular discrete setting, {\em
  shapes} refer to sharply bounded regions in which periodic patterns
appear. Despite very simple local evolution rules, very complex
structures can be generated. The well known Conway's Game of Life can
perform computations and can even emulate an universal Turing machine
(see~\cite{Wolfram} also for a historical introduction on cellular
automata).

In this Letter, we shall concentrate on a particularly simple cellular
automaton, the Abelian Sandpile Model (ASM). Originally introduced as
a statistical model of Self-Organized Criticality~\cite{BTW}, because
it shows scaling laws without any fine-tuning of an external control
parameter, it has been shown afterwards to possess remarkable
underlying algebraic properties~\cite{Dhar, Creutz}, and has been studied also
in some deterministic approaches, exactly in connection with pattern
formation~\cite{Kaplan-Ostojic-HLMPW}. The ASM has been shown to be
able to produce {\em allometry}, that is a growth uniform and constant
in all the parts of a pattern as to keep the whole shape substantially
unchanged, and thus requires some coordination and communication
between different parts~\cite{DSC}. This is at variance with
diffusion-limited aggregation and other models of growing objects
studied in physics literature so far, e.g.\ the Eden model, KPZ
deposition and invasion-percolation~\cite{KPZ-HBS}, which are
mainly models of aggregation, where growth occurs by accretion on the
surface of the object, and inner parts do not evolve significantly.

In the sandpile, the regions of a configuration periodic in space,
called {\em patches}, are the ingredients of pattern formation.
In~\cite{DSC}, a condition on the shape of patch interfaces has been
established, and proven at a coarse-grained level.

We discuss how this result is strengthened by avoiding the coarsening, and describe the
emerging fine-level structures, including linear interfaces and
rigid domain walls  with a residual one-dimensional translational
invariance. These structures, that we shall call {\em strings}, are
macroscopically extended in their periodic direction, while showing
thickness in a full range of scales between the microscopic lattice
spacing and the macroscopic volume size.

\mysection{The model}%
While the main structural properties of the ASM can be discussed on
arbitrary graphs~\cite{Dhar}, for the subject at hand here we shall
need some extra ingredients (among which a notion of translation),
that, for the sake of simplicity, suggest us to concentrate on the
original realization on the square lattice~\cite{BTW}, within a
rectangular region $\Lambda \in \mathbb{Z}^2$.

We write $i \sim j$ if $i$ and $j$ are first neighbours.  The
configurations are vectors $z \equiv \{ z_i\}_{i\in \Lambda} \in
\mathbb{N}^\Lambda$ ($z_i$ is the number of sand-grains at vertex $i$).  Let
$\bar{z}=4$, the degree of vertices in the bulk, and say that a
configuration $z$ is {\em stable} if $z_i < \bar{z}$ for all $i \in
\Lambda$. Otherwise, it is {\em unstable} on a non-empty set of sites,
and undergoes a relaxation process whose elementary steps are called
\emph{topplings}: if $i$ is unstable, we can decrease $z_i$ by
$\bar{z}$, and increase $z_j$ by one, for all $j \sim i$.  The
sequence of topplings needed to produce a stable configuration is
called an {\em avalanche}.

Avalanches always stop after a finite number of steps, which is to say
that the diffusion is strictly {\em dissipative}.  Indeed, the total
amount of sand is preserved by topplings at sites far from the
boundary of $\Lambda$, and strictly decreased by topplings at boundary
sites.
The stable configuration $\relaxx (z)$ obtained
from the relaxation of $z$,   is univocally defined,
as all valid stabilizing sequences of topplings only differ by
permutations.

We call a stable configuration \emph{recurrent} if it can be obtained through
an avalanche involving all sites in $\Lambda$, and \emph{transient}
otherwise~\cite{Dhar}.  Recurrent configurations have a
structure of Abelian group~\cite{Creutz} under the operation $z \oplus
w := \relaxx(z+w)$.
We have only a partial knowledge of the group identity for
each $\Lambda$ (see e.g.\ \cite{Creutz, LeBRossin};
recently a complete characterization has been achieved for a
simplified directed lattice, the {\em F-lattice}~\cite{noi}) 
nonetheless they are easily obtained on a computer and they provide a
first example of the intriguing complex patches in which we are
interested.  The maximally-filled configuration $z_{\rm max}$, with
$(z_{\rm max})_i = \bar{z}-1 = 3$ for all $i$, is recurrent.  More
generally, for large $\Lambda$, recurrent configurations must have
average density $\rho(z) = |\Lambda|^{-1} \sum_i z_i \geq 2 + o(1)$
(this bound is tight). So structures with density $\rho>2$, $\rho=2$
and $\rho<2$ are said respectively \emph{recurrent}, \emph{marginal}
and \emph{transient}.


A {\em patch} is a region filled with a periodic pattern. The 
{\em density} $\rho$ of a patch is the average of $z_i$ within a unit tile.
Neighbouring patches may have an interface, periodic in one dimension,
along a vector which is principal for both patches. Let us suppose that in a deterministic protocol~\cite{DSC} we
generate a region filled with polygonal patches, glued together with such a kind of
interfaces. At a vertex where $\ell \geq 3$
patches meet, label cyclically with $\alpha = 1, \ldots, \ell$ these
patches, call $\rho_\alpha$ the corresponding densities, and
$\theta_\alpha$ the angles of the interfaces between the patch
$\alpha$ and $\alpha+1$ (subscripts $\alpha=\ell+1 \equiv 1$).  These
quantities are proven to satisfy the relation in $\mathbb{Q} + i
\mathbb{Q}$
\be
\sum_{\alpha=1}^{\ell}
\left(\rho_{\alpha+1} - \rho_\alpha \right)\,
\exp (2 i \theta_\alpha) = 0 
\label{dhar}
\ee
which has non-trivial solutions only for $\ell \ge 4$~\cite{DSC}.

\begin{figure}[t]
\begin{center}
\setlength{\unitlength}{7.8pt}
\begin{picture}(24,14)
\put(0,0){\includegraphics[scale=0.65]{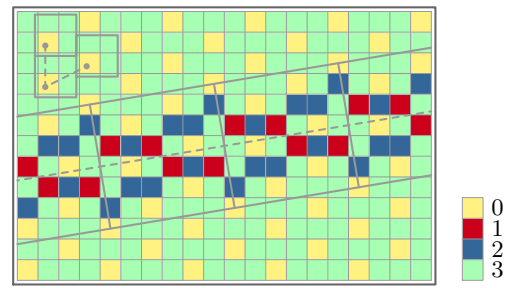}}
\put(23.5,0.65){3}
\put(23.5,1.65){2}
\put(23.5,2.65){1}
\put(23.5,3.65){0}
\end{picture}
\end{center}
\caption{\label{fig1}A string with momentum $(6,1)$, in a background
  pattern with periodicities $V = \big( (2,1),(0,2) \big)$.  String
  and background unit cells are shown in gray.  The density in the
  string tile is $\rho = (18 \cdot 3 + 8 \cdot 2 + 4 \cdot 1 + 7 \cdot
  0)/(6^2+1^2) = 2$.}
\end{figure}

A {\em string} is a one-dimensional periodic defect line, with
periodicity integer vector $\ka = (k_x, k_y)\in \mathbb{N}^2$, that we
call {\em momentum}, in a background patch, periodic in both
directions, and has $\ka$ as a periodicity vector. The background on
the two sides may possibly have a periodicity offset.  See
Fig.~\ref{fig1}.

\mysection{Results}%
In this section we report about some results obtained by our
investigations. These facts have explicit proofs, whose details will
be reported elsewhere.

\begin{figure}[b]
\begin{center}
\setlength{\unitlength}{10pt}
\begin{picture}(22.5,9.7)
\put(0,0){\includegraphics[scale=1]{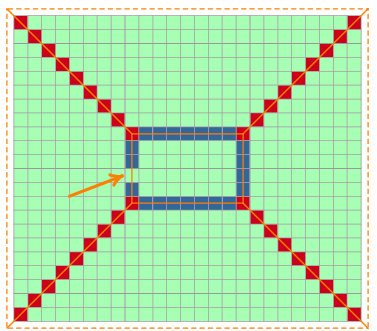}}
\put(11.8,0){\includegraphics[scale=1]{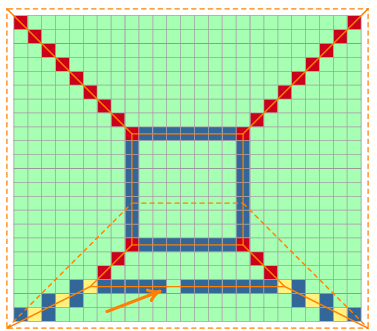}}
\end{picture}
\end{center}
\caption{\label{fig2} On the left, the configuration obtained after
  relaxation from $z_{\rm max}$ plus an extra grain of sand exactly at
  the vertex where a defect appears.  On the right, the result after
  removing the defect and the addition of one more grain.}
\end{figure}

Strings in the maximally-filled background emerge in a simple
protocol.  Consider a rectangular region $\Lambda$, and the
configuration $z_{\rm max}$.  Add one grain of sand at some vertex
$j$.  The configuration after relaxation contains an inner rectangle,
of strings $(1,0)$ and $(0,1)$, equidistant from the border of
$\Lambda$ and having $j$ on its perimeter, and its corners are
connected to the corners of $\Lambda$ with strings $(1,1)$ and
$(-1,1)$. There is one defect exactly at $j$, manifested as a single
extra grain, w.r.t.\ the underlying periodic structure. See
Fig.~\ref{fig2}. This configuration is recurrent.  Now remove this
extra grain (the configuration is now transient), and repeat the game
at some new vertex $j'$, say in the region below the inner
rectangle. In the configuration after relaxation appear also strings
$(2,1)$ and $(-2,1)$, and once again we have a defect at $j'$.  This
procedure can be iterated, and, if $\Lambda$ is large enough, strings
with higher and higher momenta generated.  Furthermore, given that the
unit tiles of momenta are classified (see the following), this
protocol is completely predictable, for arbitrary $\Lambda$, through a
purely geometric construction.

In a given recurrent background, with translation vectors 
$V = (\va_1, \va_2)$, one and only one string of momentum $\ka
= m_1 \va_1 + m_2 \va_2 = \ma V$ can be produced, if
$\gcd(m_1,m_2)=1$, and no strings of momentum $\ka$ exists for $\ka$
not of the form above.  $V$ and $V'$ are equivalent descriptions of the
background periodicity iff $V' = M V$, with $M \in \slz$.
Accordingly, $\ma'= \ma M^{-1}$. And indeed, sets of $\ma
\in \mathbb{Z}^2$ with given $\gcd$'s are the only proper subsets
invariant under the action of $\slz$.  The $\gcd$ constraint arises in
the classification of the elementary strings, because, when
$d = \gcd(k_x,k_y) > 1$, the corresponding periodic ribbon is just
constituted of $d$ parallel strings with momentum $\ka/d$.

The unit tile of each string, as well as of each patch, is symmetric
under 180-degree rotations. In particular, momenta $\ka$ and $-\ka$
describe the \emph{same} string.  The tile of a string of momentum
$\ka$ fits within a square having $\ka$ as one of the sides, so that
each string is a row of identically filled squares. This is a
non-empty statement: the tile could have required rectangular boxes of
larger aspect ratio, and even an aspect ratio depending on momentum
and background.

A string of momentum $\ka$ has an {\em energy} $E$, defined as the
difference of sand-grains, in the framing unit box of side $\ka$,
w.r.t.~$z_{\rm max}$.  We have the relation $E = |\ka|^2$, or, in
other words, the unit tile has exactly marginal density, $\rho = 2$,
irrespectively of the density of the surrounding background (as seen,
e.g., in Fig.~\ref{fig1}).

\begin{figure}[t]
\begin{center}
\setlength{\unitlength}{7.8pt}
\begin{picture}(28,15)
\put(0,0){\includegraphics[scale=0.65]{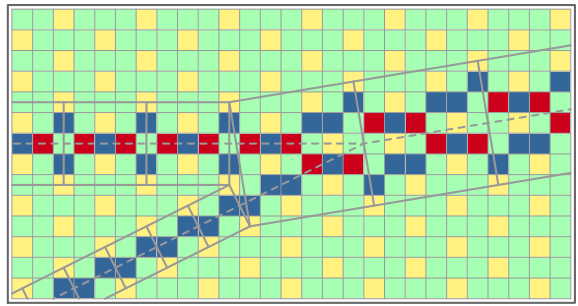}}
\end{picture}
\end{center}
\caption{\label{scattering}A scattering involving pseudo-propagators with
  momenta $(4,0)$, $(2,1)$ and $(6,1)$, on the background pattern of
  Fig.~\ref{fig1} (also symbol code is as in Fig.~\ref{fig1}).}
\end{figure}

Two strings, respectively of momentum $\pa$ and
$\qa$, can collapse in a single one of momentum $\ka$ (see
Fig.~\ref{scattering}).  In this process momentum is {\em conserved}:
$\pa + \qa= \ka$.  More precisely, the strings join together in
such a way that the square boxes surrounding the unit cells meet at an
extended {\em scattering vertex}, a triangle of sides of lengths equal
to $|\ka|$, $|\pa|$ and $|\qa|$, rotated by 90 degrees
w.r.t.~the corresponding momenta: given this geometrical construction,
momentum conservation rephrases as the oriented perimeter of the
triangle being a closed polygonal chain.

Local momentum conservation and the \hbox{$\ka \leftrightarrow -\ka$} symmetry are reminiscent of equilibrium of tensions, in a planar network of tight material strings, from which the name.

On  networks, this local conservation is extended to a global constraint. Choose an orthogonal frame $(x,t)$, and orient momenta in the direction of increasing $t$. Then, sections at fixed $t$ are all crossed by the same total momentum.
Rigid extended domain walls between periodic patterns, satisfying similar local and global conservations, appear in certain tiling models~\cite{walls}, which remarkably show a Yang-Baxter integrable structure, where the corresponding strings are usefully interpreted as world-lines of particles in the $(x,t)$-frame. Note, however, that, at variance with these models, in the ASM we have an infinite tower of excitations, for a given background, and infinitely many different backgrounds too.

In the maximally-filled background, because of the ($D_4$ dihedral) symmetry, we can restrict without loss
of generality to study strings of momentum $\ka$ with both components
positive.
For each such $\ka$ with $\gcd(k_x,k_y)=1$, simple modularity
reasonings show that there exists a {\em unique} ordered pair of
momenta $\pa$ and $\qa$, with non-negative components, such that
$\pa + \qa= \ka$ and the matrix
$\left( \begin{smallmatrix}
p_x & p_y \\ q_x & q_y \end{smallmatrix} \right)$
is in $\slz$.
We write in this case $\ka \leftarrow (\pa, \qa)$.
For example,
$(10,3) \leftarrow \big( (7,2),(3,1) \big)$. 
The endpoint of $\pa$, starting from the top-left corner of the $\ka$ framing box, 
is the (unique) lattice point which is nearest to the top-side of the box. 
This alternative definition generalizes to non-trivial backgrounds, and the $m_1 \va_1 + m_2 \va_2$ sublattice.



We have found a simple algorithm to derive the string textures.
The tile
for $\ka \leftarrow (\pa,\qa)$ is essentially composed of four interlaced tiles, two for
$\pa$ and two for $\qa$, adjacent to the four vertices of the $\ka$ box.
This opens a problem of consistency for the
overlapping region, of side $\sim |\pa-\qa|$, which is solved by the fact that, for $\ka \leftarrow (\pa, \qa)$,
then $\pa \leftarrow (\pa-\qa, \qa)$ if $|\pa|>|\qa|$, 
and $\qa \leftarrow (\pa, \qa-\pa)$ otherwise,
again by a property of $\slz$
(the reader can see in
Fig.~\ref{scattering} the similarity of the tile of a composed string
with the ones of its component).
By iterating this
procedure, starting with the strings of minimum momentum, we can
understand the textures of the full catalog of strings, and, in
particular, this makes the protocol for producing networks of strings completely predictable.

Let us go back to the problem of $\ell$ interfaces which meet at a
given corner, but allow now, besides interfaces between patches,
incident strings. Following the analysis of~\cite{DSC}, we introduce the
graph-vector $T = \{ T_i \}$, where $T_i$ is the number of topplings
at $i$ in the relaxation of the starting configuration, and study its
characteristics in a region that, in the starting configuration, was
uniformly filled with a patch.  However, now we allow for toppling
distributions which are piecewise both quadratic and linear (the
linear term was neglected in~\cite{DSC}, as subleading in the
coarsening).

For any relevant direction $\alpha$, allow for a patch interface, or a
string, or both.
Call $\tilde{E}^{(\alpha)}$ the difference for \emph{unit length}
(not for period), in the total
number of grains of sand w.r.t.~$z_{\rm max}$, due to presence of a
string, i.e.\ $\tilde{E}^{(\alpha)} = E^{(\alpha)}/|\ka^{(\alpha)}|$,
or the contribution from a non-zero impact parameter in the interface.
It can be shown, by reasonings as in~\cite{DSC}, that 
the difference between the extrapolated toppling profile for two contiguous
patches, at a polar coordinate $(r, \theta)$,
must be of the form
\be
\begin{split}
\label{diff} 
T^{(\alpha+1)}_{r, \theta}
 - T^{(\alpha)}_{r, \theta}
&= \frac{r^2}{2} \left(\rho_{\alpha+1}
- \rho_\alpha\right)\,\sin^2(\theta - \theta_\alpha)
\\
& \quad + \,r\, \tilde{E}^{(\alpha)} \sin(\theta - \theta_\alpha)
+ \mathcal{O}(1)\, .
\end{split}
\ee
Then, by summing over $\alpha$ and matching separately the quadratic
and linear terms, we conclude that, for each $\theta$,
\begin{gather}
\left\{
\begin{array}{l}
\rule{0pt}{6pt}%
\sum_{\alpha=1}^{\ell} \left(\rho_{\alpha+1} -
\rho_\alpha\right)\,\sin^2(\theta - \theta_\alpha) = 0
\\
\rule{0pt}{16pt}%
\sum_{\alpha=1}^{\ell}  \tilde{E}^{(\alpha)} \sin (\theta -
\theta_\alpha) = 0
\end{array}
\right.
\end{gather}
so that, besides the anticipated
equation~(\ref{dhar}) for patches alone, that
was deduced in~\cite{DSC}, we obtain
\begin{gather}
\label{eq.65458b}
\sum_{\alpha=1}^{\ell}  \tilde{E}^{(\alpha)} \,\exp( i \theta_\alpha) = 0
\end{gather}
which describes the string and interface-offset contributions. 

In (\ref{dhar}), the first non-trivial value for $\ell$ is
4~\cite{DSC}.  In our generalization, 4 is the minimal value for the
number of patches \emph{plus} the number of strings, and thus includes
new possibilities: a \emph{scattering} event, with three incident
strings in a single background, as in Fig.~\ref{scattering}, and the case of two strings and two
patches, producing diagrams reminiscent of total \emph{reflection} and
\emph{refraction} in optics, so that the specialization of
(\ref{eq.65458b}) can be read as a \emph{Snell's law} for ASM strings.
For the case of three strings on a common background  $\cal B$, we get
\be
\sum_{\alpha=1}^{\ell}
\frac{E^{(\alpha)}}{|\ka^{(\alpha)}|^2} \, \ka^{(\alpha)} = 0\, 
\ee
which shows that momentum conservation implies a dispersion relation
 of the form
$E^{(\alpha)} = c_{\cal B} |\ka^{(\alpha)}|^2$,
and viceversa.

\begin{figure}[t]
\begin{center}
\setlength{\unitlength}{7.5pt}
\begin{picture}(32,11)
\put(0,0){\includegraphics[scale=0.75]{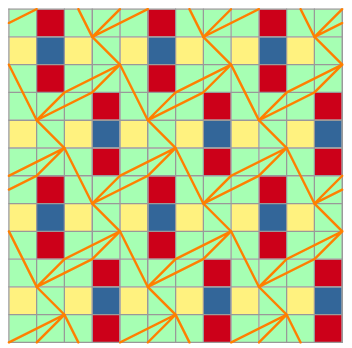}}
\put(11,0){\includegraphics[scale=0.75]{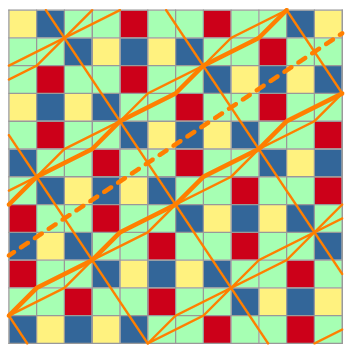}}
\put(22,0){\includegraphics[scale=0.75]{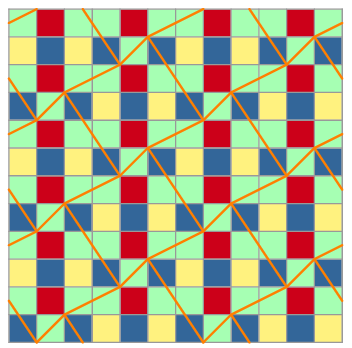}}
\end{picture}
\end{center}
\caption{\label{fig_patch}The recurrent, marginal and transient
  patches constructed from the propagator $\ka=(3,2) \leftarrow \big(
  (2,1), (1,1) \big)$ in the $z_{\rm max}$ background, having
  densities $\rho=2$ and $2 \pm 1/12$ (symbol code is as in
  Fig.~\ref{fig1}).}
\end{figure}

The classification of the strings preludes to a classification of
the patches. To any string of momentum $\ka$, univocally decomposed as
$\ka \leftarrow (\pa, \qa)$,
we can associate three patches, respectively recurrent, marginal and
transient, through a geometrical construction, involving $\pa$ and
$\qa$, sketched in an example in Fig.~\ref{fig_patch}.

Reflection and refraction events also appear, if $\ka \leftarrow (\pa,
\qa)$, in a single string of momentum $\ka' = m \pa + \qa$ (for $m$ a
large integer), and in the scattering of $m \pa + \qa$ into $\qa$ and
$m$ parallel $\pa$ strings. Indeed, as a consequence of the recursive
construction of the string textures, the string of momentum $\ka'$ is
forced to look as a strip-shaped patch of $m$-period width, of the
marginal tile associated to $\pa$, crossed by a `soft' string, that
reflects twice per period $\ka'$, up to ultimately leaving the marginal
patch, through a refraction,  and propagates in the recurrent background.

We shall show elsewhere that the interplay between strings and
patches, both at the level of classification, and of evolution in
deterministic protocols, is the key-ingredient  to clarify
allometry in pattern formation for the ASM, and to design
new protocols in which short-scale defects are
totally absent. The resulting structure is a fractal,  a
Sierpi\'nski triangoloid, where the theoretical formula~(\ref{dhar}) has
infinitely many distinct realizations.



\begin{thebibliography}{99}

\bibitem{Thompson}
D.W.~Thompson,
{\it On growth and form,}
Dover reprint of 1942 2nd ed., 1992 (1st ed., 1917).

\bibitem{Thom}
R.~Thom,
{\it Structural stability and morphogenesis,}
W.A.\ Benjam, 1972.

\bibitem{DLA}
T.A.~Witten Jr, L.M.~Sander, 
Phys.\ Rev.\ Lett.\ 
 {\bf 47}, 1400 (1981)
and Phys.\ Rev.\ B{\bf 27}, 5686 (1982).
P.~Meakin, Phys. Rev. A{\bf 27}, 1495 (1983). 
T.~Vicsek, 
{\it Fractal growth phenomena,} World Scientific, Singapore, 1989.

\bibitem{Mandelbrot}
B.~Mandelbrot, 
{\it The Fractal Geometry of Nature,}
W.H.\ Freeman, New York, 1982.

\bibitem{CD}
B.~Chopard, M.~Droz,
{\it Cellular Automata modeling of physical systems,}
Cambridge Univ.\ Press, 1998.


\bibitem{Wolfram}
S.~Wolfram,
{\it A new kind of Science,}
Wolfram Med., 2002.

\bibitem{BTW}
P.~Bak, C.~Tang, K.~Wiesenfeld,
Phys.\ Rev.\ Lett.\ 
{\bf 59}, 381 
(1987).

\bibitem{Dhar}
D.~Dhar,
Phys.\ Rev.\ Lett.\ {\bf 64}, 1613 (1990).
S.N.~Majumdar, D.~Dhar,
Physica A{\bf 185}, 129 (1992).
D.~Dhar,
Physica A{\bf 263}, 4 (1999).

\bibitem{Creutz}
M.~Creutz, 
Comp.\ Phys.\ {\bf 5}, 198 (1991).

\bibitem{Kaplan-Ostojic-HLMPW}
S.H.~Liu, T.~Kaplan, L.J.~Gray,
Phys.\ Rev.\ A{\bf 42}, 3207
(1990).
S.~Ostojic,
Physica A{\bf 318}, 187 
(2003).
A.~Holroyd, L.~Levine, K.~M\'esz\'aros, Y.~Peres, J.~Propp, D.~Wilson,
Progress in Prob.\ {\bf 60}, 331 (2008).

\bibitem{DSC}
D.~Dhar, T.~Sadhu, S.~Chandra,
 Europhys.\ Lett.\ 
{\bf 85}, 48002 (2009).

\bibitem{KPZ-HBS}
M.~Kardar, G.~Parisi, Y.-C.~Zhang, 
 Phys.\ Rev.\ Lett.\  {\bf 56}, 889 
(1986).
H.J.~Herrmann, Phys.\ Rep.\ {\bf 136}, 153 (1986). 
L.~Barabasi, H.E.~Stanley, 
{\it Fractal concepts in surface growth,}
Cambridge Univ.\ Press, 1995.



\bibitem{LeBRossin}
Y.~Le\,Borgne, D.~Rossin,
Discr.\ Math.\ {\bf 256}, 775 (2002).

\bibitem{noi}
S.~Caracciolo, G.~Paoletti, A.~Sportiello, 
J.\ Phys.\ A: Math.\ Theor.\ {\bf 41}, 495003 (2008).

\bibitem{walls}
M.~Widom, Phys.\ Rev.\ Lett.\ {\bf 70}, 2094 (1993).
P.A.~Kalugin, J.\ Phys.\ A: Math.\ Gen.\ {\bf 27}, 3599 (1994).
A.~Verberkmoes, B.~Nienhuis, Phys.\ Rev.\ Lett.\ {\bf 83}, 3986 (1999).

\end{thebibliography}
\end{document}